\documentclass[aps,prl, twocolumn,floatfix,superscriptaddress,preprintnumbers,
               tightenlines,showpacs,showkeys,nofootinbib,notoccite]{revtex4-1}
\usepackage[utf8]{inputenc}
\usepackage[colorlinks=true,citecolor=blue,linkcolor=blue]{hyperref}
\usepackage[normalem]{ulem}
\usepackage{amsmath,amssymb, mathrsfs}
\usepackage{epsfig}
\usepackage{graphicx}               
\usepackage{url}
\usepackage{color}
\usepackage{slashed}
\usepackage{multirow}
\usepackage{placeins}
\usepackage[dvipsnames]{xcolor}
\usepackage{epstopdf}
\usepackage{soul}
\usepackage{tikz}
\usepackage[capitalise, english]{cleveref}
\usepackage{siunitx}
\usepackage{xspace}
\usepackage{booktabs}
\usetikzlibrary{trees}
\usetikzlibrary{decorations.pathmorphing}
\usetikzlibrary{decorations.markings}

\newcommand\myshade{80}
\colorlet{mylinkcolor}{ForestGreen}
\colorlet{mycitecolor}{Red}
\colorlet{myurlcolor}{violet}

\hypersetup{
  linkcolor  = mylinkcolor!\myshade!black,
  citecolor  = mycitecolor!\myshade!black,
  urlcolor   = myurlcolor!\myshade!black,
  colorlinks = true
}

\definecolor{jblue}{RGB}{20,50,100}
\definecolor{npurple}{RGB} {153, 51, 204}
\definecolor{wred}{RGB}{217,0,56}
\definecolor{white}{RGB}{255,255,255}

\definecolor{korange}{RGB}{235, 80,  43}
\definecolor{korange2}{RGB}{245, 100,  63}
\definecolor{kyelloworange}{RGB}{255, 210,  110}
\definecolor{kyelloworange2}{RGB}{240, 170,  90}
\definecolor{kred}{RGB}{204,  102, 153}
\definecolor{kpurple}{RGB}{153,  61, 190}
\definecolor{kpurplelight}{RGB}{213,  161, 230}

 \definecolor{tobycolour}{rgb}{.5,.0,.5}

\DeclareSIUnit\year{yr}
\DeclareSIUnit\pc{pc}
\DeclareSIUnit\ergs{ergs}
\DeclareSIUnit\msun{\ensuremath{M_\odot}}
\allowdisplaybreaks

\newcommand{\ra}[1]{\renewcommand{\arraystretch}{#1}}            

\makeatletter
\providecommand*{\diff}%
  {\@ifnextchar^{\DIfF}{\DIfF^{}}}
\def\DIfF^#1{%
  \mathop{\mathrm{\mathstrut d}}%
    \nolimits^{#1}\gobblespace}
\def\gobblespace{%
  \futurelet\diffarg\opspace}
\def\opspace{%
  \let\DiffSpace\!%
  \ifx\diffarg(%
    \let\DiffSpace\relax
  \else
    \ifx\diffarg[%
      \let\DiffSpace\relax
    \else
        \ifx\diffarg\{%
        \let\DiffSpace\relax
      \fi\fi\fi\DiffSpace}

\usepackage{tikz,xcolor,hyperref}

\definecolor{lime}{HTML}{A6CE39}
\DeclareRobustCommand{\orcidicon}{\hspace{-1mm}
	\begin{tikzpicture}
	\draw[lime, fill=lime] (0,0) 
	circle [radius=0.16] 
	node[white] {{\fontfamily{qag}\selectfont \tiny \,ID}};
	\draw[white, fill=white] (-0.0525,0.095) 
	circle [radius=0.007];
	\end{tikzpicture}
	\hspace{-3mm}
}

\foreach \x in {A, ..., Z}{\expandafter\xdef\csname orcid\x\endcsname{\noexpand\href{https://orcid.org/\csname orcidauthor\x\endcsname}
			{\noexpand\orcidicon}}
}


\begin{document}

\title{Low Mass Black Holes from Dark Core Collapse}

\author{Basudeb Dasgupta\orcidA{}}
\email{bdasgupta@theory.tifr.res.in}
\affiliation{Tata Institute of Fundamental Research, Homi Bhabha
	Road, Mumbai 400005, India}

\author{Ranjan Laha\orcidB{}} 
\email{ranjan.laha@cern.ch}
\affiliation{Theoretical Physics Department, CERN, 1211 Geneva, Switzerland}
\affiliation{Centre for High Energy Physics, Indian Institute of Science, C.\,V.\,Raman Avenue, Bengaluru 560012, India}

\author{Anupam Ray\orcidC{}}
\email{anupam.ray@theory.tifr.res.in}
\affiliation{Tata Institute of Fundamental Research, Homi Bhabha
	Road, Mumbai 400005, India}

\date{\today}


\begin{abstract}
Unusual masses of  black holes being discovered by gravitational wave experiments pose fundamental questions about the origin of these black holes. Black holes with masses smaller than the Chandrasekhar limit $\approx1.4\,M_\odot$ are essentially impossible to produce through stellar evolution. We propose a new channel for production of low mass black holes: stellar objects catastrophically accrete non-annihilating dark matter, and the small dark core subsequently collapses, eating up the host star and transmuting it into a black hole. The wide range of allowed dark matter masses allows a smaller effective Chandrasekhar limit, and thus smaller mass black holes. We point out several avenues to test our proposal, focusing on the redshift dependence of the merger rate. We show that redshift dependence of the merger rate can be used as a probe of the transmuted origin of low mass black holes.
\end{abstract}

\maketitle
\preprint{TIFR/TH/20-32, CERN-TH-2020-145}

\emph{Introduction --} 
The recent detections of GW190425\,\cite{Abbott:2020uma} and GW190814\,\cite{Abbott:2020khf}, which are either the heaviest neutron stars (NSs) or the lightest black holes (BHs) ever seen, have ignited interest in $\mathcal{O}$(1)\,$M_\odot$ BHs\,\cite{Flitter:2020bky, Kinugawa:2020tbg, Clesse:2020ghq, Jedamzik:2020omx, Vattis:2020iuz}. Usual stellar evolution cannot lead to sub-Chandrasekhar mass BHs, and the observation of such black holes would augur new physics. With the remarkable advances in gravitational wave (GW) and multi-messenger astronomy, the detection of a sub-Chandrasekhar mass ($\lesssim\,$1.4\,$M_\odot$) BH may be just around the corner.
 
The key question, assuming a future GW observation of a merger involving a sub-solar-mass object, is how to pinpoint its identity? One recently popular candidate is a primordial black hole (PBH). It is a well-motivated dark matter (DM) candidate\,\cite{Zeldovich:1967lct,Hawking:1971ei,Carr:1974nx,Chapline:1975ojl} that can constitute all of the cosmological DM\,\cite{Katz:2018zrn,Montero-Camacho:2019jte,Smyth:2019whb}. After a critical reappraisal~\mbox{\cite{Abbott:2016blz,Bird:2016dcv,Clesse:2016vqa,Sasaki:2016jop}}, numerous constraints now exist on their abundance\,(\cite{Arbey:2019vqx, Clark:2016nst, Poulter:2019ooo, Boudaud:2018hqb, DeRocco:2019fjq, Laha:2019ssq, Dasgupta:2019cae, Laha:2020ivk, Allsman:2000kg, Tisserand:2006zx, Niikura:2019kqi, Croon:2020ouk, Oguri:2017ock, Zumalacarregui:2017qqd, Authors:2019qbw, Nitz:2020bdb,Kavanagh:2018ggo, Brandt:2016aco, Monroy-Rodriguez:2014ula, Serpico:2020ehh, Hektor:2018qqw, Manshanden:2018tze, Hektor:2018rul, Sammons:2020kyk, Lu:2020bmd,Kim:2020ngi,Laha:2020vhg}, see also \,\cite{Carr:2020gox,Carr:2020xqk,Green:2020jor} and references therein). However, there is no mechanism that naturally produces the correct abundance of PBHs. The initial abundance of PBHs is exponentially sensitive to the spectrum of density perturbations and the threshold for collapse; fine-tuning of parameters is required to achieve the observed DM abundance~\cite{Aghanim:2018eyx}. 

An alternative possibility is that the low mass BHs are of a non-primordial nature. DM accretion in a star followed by its dark core collapse~\cite{Kouvaris:2018wnh} or the exotic cooling of an atomic dark matter cloud~\cite{Shandera:2018xkn} can lead to low mass BHs. However, previously proposed models employing a fermionic asymmetric DM with non-negligible self-interaction or dark quantum electrodynamics are not generic. Transit of a tiny PBH through a compact object and subsequent conversion of the host to a BH is also thought to be a novel way to produce sub-Chandrasekhar and $\mathcal{O}$(1)\,$M_\odot$ BHs\,\cite{Capela:2013yf,Takhistov:2017bpt}. However, the estimated capture rate of a tiny PBH by a NS was recently shown to be quite small, $\sim 10^{-17}$\,yr$^{-1}$ for a NS residing in a Milky-Way-like galaxy with ambient DM density, $\rho_{\chi} =1$\,GeV\,cm$^{-3}$\,\cite{Montero-Camacho:2019jte,Genolini:2020ejw}.  (We use $\hbar=c=1$ units hereafter.) The capture rate scales linearly with the ambient DM density and has a strong dependence on the velocity dispersion, $(\bar{v}^{-3})$, so an ${\mathcal O} (1)$ Gyr old NS in a DM dense region ($\rho_{\chi} =10^3$ GeV cm$^{-3}$) inside a globular cluster $(\bar{v} \sim 10^{-5})$ can, in principle, implode due to a PBH transit. However, such overdense DM cores in a globular cluster are quite speculative and not yet well established. It has, in fact, been shown that globular clusters do not have any DM over-densities\,\cite{2011ApJ...741...72C, Naoz:2014bqa, Hurst:2019okj}. Hence, the explanation of a sub-Chandrasekhar or $\mathcal{O}$(1)\,$M_\odot$ BH due to a PBH transit hinges on the contentious assumption of a high DM density in globular clusters, and remains uncertain until the provenance of globular clusters is settled.

In this \emph{Letter}, we point out a simple mechanism that transmutes a sub-Chandrasekhar or $\mathcal{O}$(1)\,$M_\odot$ star to a comparable ``low mass BH'', and propose several tests for the proposal focusing on the cosmic evolution of the merger rates. DM candidates share the universal feature that they have a mass and interactions with ordinary matter. We show, without appealing to any other exotic features, that this is sufficient for making transmuted black holes (TBHs). Continued accumulation of non-annihilating DM particles in the core, followed by their gravitational collapse at a modified Chandrasekhar limit set by DM mass and spin, can produce sub-Chandrasekhar or $\mathcal{O}$(1)\,$M_\odot$ TBHs that are a viable alternative to PBHs. We answer a few basic questions: what particle physics parameter space can they explore, how to test their origin, and especially, how to distinguish them from PBHs?

\begin{figure*}[t]
	 \includegraphics[width=1.0\textwidth]{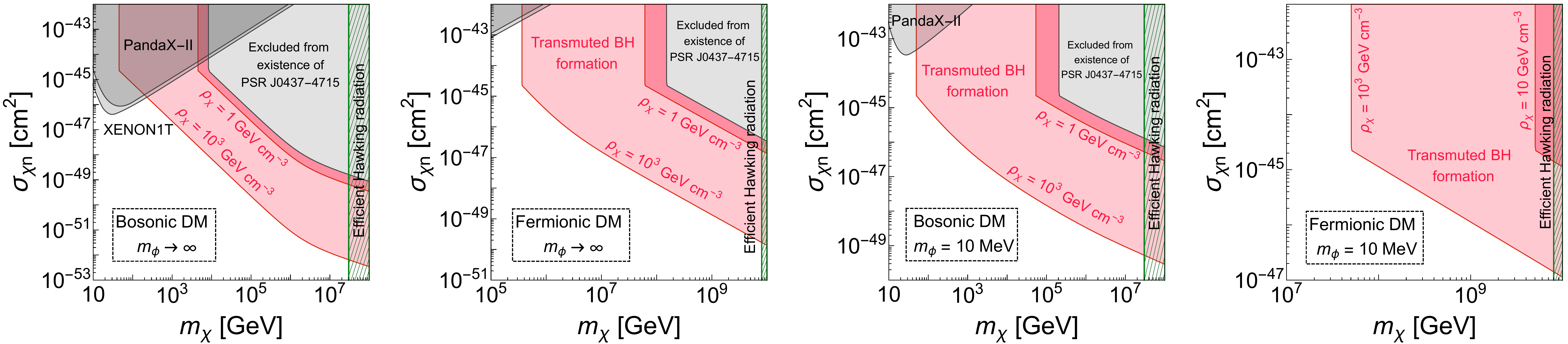}
	\caption[]{DM mass and scattering cross section required for a dark core collapse and subsequent transmutation of a 1.3\,$M_{\odot}$ NS to a comparable mass BH are shown in the red shaded regions.  The two panels on the left (respectively, right) correspond to interaction between DM and stellar nuclei mediated by an infinitely heavy mediator, i.e., $m_{\phi} \gg$ recoil momentum (resp.\,10 MeV scalar). Scenarios with non-annihilating bosonic or fermionic DM are marked.  Two representative values of ambient DM density, $\rho_{\chi}$\,=\,1~and~$10^3$\,GeV\,cm$^{-3}$ (only for the right panel, $\rho_{\chi}$\,=\,10~and~$10^3$\,GeV\,cm$^{-3}$), are considered.  Exclusion limits from underground direct detection experiments {\sc PandaX-II}\,\cite{Ren:2018gyx} and {\sc XENON1T}\,\cite{Aprile:2018dbl} as well as from existence of an $\sim$ 7 Gyr  old nearby pulsar PSR J0437-4715\,\cite{Manchester:2004bp,McDermott:2011jp,Garani:2018kkd,Dasgupta:2020dik} are also shown by the gray shaded regions. Green hatched regions mark the parameter space where efficient Hawking evaporation stops the implosion of the NS. The region of no thermalization is many orders of magnitude below\,\cite{Garani:2018kkd}, and is not shown for clarity.}
	\label{fig: particle physics constraints}
\end{figure*}

\emph{Methods \& Results --} Transmutation of stellar objects to BHs due to core collapse has been studied extensively, as a tool to set constraints on DM-nucleon scattering cross section from the existence of old NSs\,\mbox{\cite{Goldman:1989nd,deLavallaz:2010wp,Kouvaris:2011fi,McDermott:2011jp,Bell:2013xk,Bramante:2013hn,Bramante:2013nma,Garani:2018kkd,Dasgupta:2020dik}}, from connection with type-Ia supernovae\,\cite{Janish:2019nkk, Graham:2018efk, Bramante:2015cua, Acevedo:2019gre}, as well as connections to several other astrophysical phenomenon\,\cite{Bramante:2014zca,Bramante:2016mzo,Fuller:2014rza,Bramante:2017ulk,East:2019dxt}. Non-annihilating DM~\cite{Zurek:2013wia,Petraki:2013wwa} scatters with stellar nuclei, gets captured via single\,\cite{Press:1985ug,gould1} or multiple scattering\,\cite{Bramante:2017xlb, Ilie:2020vec, Dasgupta:2019juq}, and accumulates inside a stellar object linearly with time.  A precise estimate of the total number of captured DM particles inside a stellar object can be found in\,\cite{Dasgupta:2020dik, Bell:2020jou}  for interactions mediated by any arbitrary mass mediators, and in the contact interaction approximation, respectively. Transmutation occurs when the captured DM particles satisfy the collapse criterion, i.e., $N_{\chi}\rvert_{t_{\textrm{age}}} \geq \text{max} \left[N^{\text{Cha}}_{\chi} , N^{\text{self}}_{\chi} \right]$, where $N_{\chi}\rvert_{t_{\textrm{age}}}$ is the total number of DM particles accumulated within a celestial object throughout its age $t_{\textrm{age}}$.  $N^{\text{Cha}}_{\chi}$ and $N^{\text{self}}_{\chi}$, respectively, denote the Chandrasekhar limit (which depends on the DM particle spin) and the number of DM particles required for initiating the self-gravitating collapse. For bosonic (fermionic) DM, zero point energy is provided by the Heisenberg uncertainty (Pauli exclusion).  The Chandrasekhar limit, $N^{\text{Cha}}_{\chi}$, for bosonic DM, $ \sim 1.5 \times 10^{34} \left( 100 \,\text{GeV}/m_{\chi} \right)^2$ can be met more easily than for its fermionic counterpart, $ \sim 1.8 \times 10^{51} \left( 100 \,\text{GeV}/m_{\chi} \right)^3$, explaining an easier transmutation for bosonic DM\,\cite{McDermott:2011jp,Garani:2018kkd}.  The required number of DM particles for self-gravitation, $N^{\text{self}}_{\chi}$, does not depend on the spin statistics of the DM particles, and is set by the condition that DM density has to exceed the baryonic density within the stellar core~\cite{McDermott:2011jp}.
  
Once the number of captured DM particles satisfies the collapse criterion, dark core collapse can ensue and a tiny BH is formed within the stellar object. This BH accumulates matter from the host star and transmutes the star into a comparable mass BH. For typical NS parameters, if this tiny BH is lighter than $\sim 10^{-20}\,{M}_{\odot}$, it evaporates faster than its mass accretion rate and cannot transmute the NS to a BH\,\cite{Kouvaris:2011fi,Dasgupta:2020dik}. For non-annihilating bosonic and fermionic DM, transmutation of a typical NS ceases due to efficient Hawking evaporation for masses $\gtrsim \mathcal O (10^7)$ and $\gtrsim \mathcal O (10^{10})$ GeV, respectively.

\begin{figure*}[t]
  	\centering
  	 \includegraphics[width=0.9\textwidth]{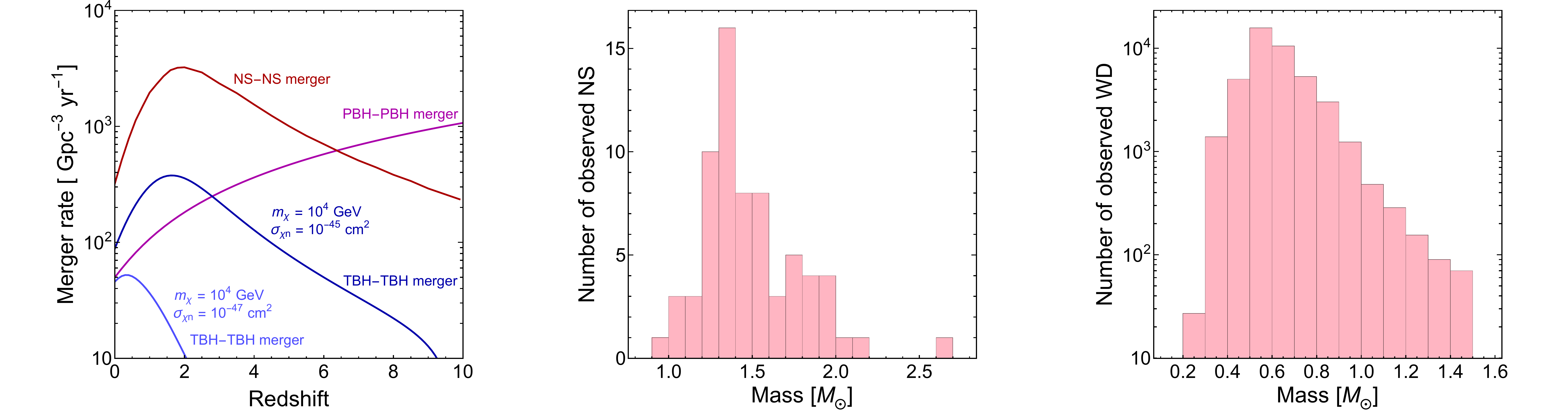}
  	\caption[]{Cosmic evolution of the binary merger rates and mass distributions of the compact objects provide a simple yet novel technique to determine the stellar or primordial origin of BHs. The left panel corresponds to the cosmic evolution of the binary PBH, NS, and TBH merger rates. For the binary NS and TBH merger rate, cosmic star formation rate is adopted from~\cite{Madau:2014bja} and they are normalized to the recent LIGO-VIRGO measurement~\cite{Abbott:2020gyp}. Non-annihilating bosonic DM with mass of 10 TeV and DM-nucleon scattering cross section of $10^{-45}$ and $10^{-47}$ $\textrm{cm}^2$ in the contact approximation are assumed for the estimation of binary TBH merger rate. The PBH merger rate is estimated by considering 1.3 $M_{\odot} -$ 1.3 $M_{\odot}$ PBH binary and a dark matter fraction $f_{\rm PBH}=10^{-3}$, which enters as $f_{\rm PBH}^{53/37}$~\cite{Raidal:2018bbj,Chen:2018czv}. The middle (right) panel corresponds to the mass distribution of all observed NSs (WDs)\,\cite{Ozel:2016oaf,2018MNRAS.480.4505J}. Mass distributions of the progenitors can be compared against some well-motivated PBH mass distributions to examine the origin of low mass BHs.}
  	\label{fig: mass distribution merger rate}
  \end{figure*}

Fig.\,\ref{fig: particle physics constraints} shows the DM parameter space where a NS (with mass $1.3\,{M}_{\odot}$) can transmute to a low mass BH for either bosonic or fermionic DM, for two choices of ambient DM density.  DM-nucleon interactions mediated by an infinitely heavy mediator (light mediator of mass 10 MeV) is assumed in the first (last) two panels. In the contact interaction approximation, asymmetric bosonic (fermionic) DM of mass $\mathcal O$(100) GeV ($\mathcal O$(1) PeV) in a DM dense environment can lead to a sub-Chandrasekhar mass BH. For DM-nucleon interaction mediated via lighter mediators, transmutation of compact objects is easier, as exclusion limits weaken and implosions can be achieved with a wider range of parameters. Similar analysis can also be performed for a solar mass white dwarf (WD). However, because of the lower baryonic density compared to a NS, the implosion criterion is harder to achieve for a WD.  For implosion of a solar mass WD with ambient DM density $10^3$ GeV cm$^{-3}$, the scattering cross section has to be $\gtrsim 10^{-44}$ cm$^2$ for a 10 PeV asymmetric bosonic DM, whereas, the corresponding cross-section for a NS with the same ambient DM density is $\sim 10^{-48}$ cm$^2$. The transmutation does not happen for small DM masses, causing the cutoff on the left side of the red shaded region in Fig.\,\ref{fig: particle physics constraints}. This is because the number of DM particles required for initiating self-gravitation ($N^{\rm{self}}_{\chi}$), as well as the Chandrasekhar limit ($N^{\rm{cha}}_{\chi}$), increases for lighter DM $[N^{\rm{self}}_{\chi} \sim 1/m^{5/2}_{\chi}$, $N^{\rm{cha}}_{\chi} \sim 1/m^{2}_{\chi} \left(1/m^{3}_{\chi}\right)$ for bosonic (fermionic) DM] and the number of captured DM particles is not sufficient to satisfy the dark core collapse criterion. Note that, we do not consider the possibility of  Bose-Einstein condensate (BEC) formation because the temperature required for BEC formation, $T_{\rm{BEC}}\approx ({2 \pi}/m_\chi)\left({n_{\chi}}/{2.612}\right)^{2/3}$\,\cite{McDermott:2011jp} is almost always less than the core temperature of the stellar object ($T = 2.1\times 10^6$\,K) that we consider.
   
How would one detect these low mass BHs and could one distinguish them from other compact objects? We now attempt to address these two questions.  In the left panel of Fig.\,\ref{fig: mass distribution merger rate} we show the redshift evolution of the binary merger rates for PBHs, NSs, and TBHs (for two representative choices of DM mass $m_\chi$ and DM-nucleon scattering cross section $\sigma_{\chi n}$). The merger rate of PBH binaries keeps rising with higher redshift as the PBH binaries can form efficiently in the early Universe~\cite{Sasaki:2018dmp,Sasaki:2016jop,Ali-Haimoud:2017rtz,Chen:2018czv,Raidal:2018bbj}. On the other hand, the merger rate of binary NSs, $R_{\rm{NS}} (t)$\,\cite{Taylor:2012db} traces the cosmic star formation rate\,\cite{Madau:2014bja,Porciani:2000ag}
  \begin{align}
  R_{\rm{NS}} (t) = \int_{t_f=t_*}^{t} dt_f \frac{dP_m}{dt}(t-t_f)\lambda\frac{d\rho_*}{dt}(t_f)\,,
  \label{eq:NS}
  \end{align}
and peaks at an $\mathcal O$(1) redshift when the star formation rate is highest. In Eq.\,(\ref{eq:NS}), $\frac{d\rho_*}{dt} (t_f)$ denotes the cosmic star formation rate at binary formation time $t_f$, $\lambda$ is the number of coalescing NS binaries per unit star forming mass, and $\frac{dP_m}{dt} (t-t_f)$ is the probability density distribution of binary NSs merging within the time interval $(t-t_f)$ after formation. For this analysis, we have used $\lambda = 10^{-5} M^{-1}_{\odot}$, earliest star formation time $(t_*)=4.9 \times 10^8$ year which corresponds to $z=10$, $\frac{dP_m}{dt} \propto (t-t_f)^{-1}$\,\cite{Taylor:2012db}, and adopted the cosmic star formation rate from~\cite{Madau:2014bja}. The merger rate of TBH binaries, $R_{\rm{TBH}} (t) $, is systematically lower than of NS binaries, $R_{\rm{NS}} (t)$, as only a fraction of them implode depending on the time required for transmutation $(\tau_{\rm{trans}})$, and depends on the NS population in the galaxies as well as evolution of the DM density in the galaxies,
 \begin{align}
  R_{\rm{TBH}} (t) = \sum_{i} f_i \int_{t_f=t_*}^{t} dt_f \frac{dP_m}{dt}(t-t_f) \lambda\frac{d\rho_*}{dt}(t_f) \times \nonumber\\   \Theta\left\{t-t_f-\tau_{\rm{trans}}\left [m_\chi,\sigma_{\chi n},\rho_{\textrm{ext},i}(t)\right] \right\}\,.
  \label{eq:TBH}
  \end{align} 
In Eq.\,(\ref{eq:TBH}), we assume that NS binaries live in Milky-Way-like galaxies and are distributed uniformly in $r=(0.01,0.1)$ kpc, where $r$ denotes the distance from the  Galactic Center. Binning $r$ into ${K}$ bins and denoting the fraction of NS binaries in the $i^{\textrm{th}}$ bin as $f_i$ one has $f_i=1/K$. We further assume that $f_i$ do not evolve with time, but the ambient DM density in the $i^{\textrm{th}}$  bin, $\rho_{\textrm{ext},i}$, does evolve with time. We assume that the DM density in each halo is given by the Navarro-Frank-White profile $\rho_{\rm ext}(r)=\rho_s/\left[(r/r_s)(1+r/r_s)^2\right]$, such that the average density within the virial radius $r_{\rm vir}$ is $200\rho_c(z)$. The parameters $\rho_s$ and $r_s$ are expressible in terms of the critical density of the Universe $\rho_c(z)$, the concentration parameter $c_{200}=r_{\rm vir}/r_s$, and the mass inside the virial radius $M_{200}$. For Milky Way like galaxies we take $c_{200}=13.31$ and $M_{200}=0.82 \times 10^{12}\,M_\odot$\,\cite{2020MNRAS.494.4291C}, so the time evolution of $\rho_{{\rm ext},i}$ is determined by evolution of $\rho_c(z)=\rho_c(0) \left[ \Omega_{\Lambda}+\Omega_m(1+z)^3\right]$, where $\Omega_m, \Omega_{\Lambda}$ are the present day density parameters for matter and dark energy, respectively~\cite{Aghanim:2018eyx}. From the expression for the merger rate, it is evident that the rate of TBH binary mergers decreases with increase in $\tau_{\rm{trans}}$. Therefore, for a fixed DM mass and DM-nucleon scattering cross section, the merger rate of TBH binaries decreases with higher redshift because NS binaries at higher redshift do not have the time for DM accretion required for implosion. Of course, given a DM mass, decrease in DM-nucleon scattering cross section leads to higher $\tau_{\rm{trans}}$, and, hence, lower merger rate. 

This distinct redshift dependence of the merger rates, especially at higher redshifts, can be measured with the upcoming third generation  GW detectors like {\sc Cosmic Explorer}~\cite{Koushiappas:2017kqm}, {\sc Einstein Telescope}\,(ET)~\cite{Evans:2016mbw} and space-based GW detector {\sc Pre-DECIGO}\,\cite{Nakamura:2016hna} which will distinguish the transmutation via implosion scenario from PBHs. In Table \,\ref{table}, we estimate the possible detection rates of TBH binary mergers. The expected detection rate, $N_D$, is simply given by~\cite{Taylor:2012db,Taylor:2011fs}
 \begin{multline}
 N_D=\int_{z=0}^{\infty} dz \frac{4\pi D^2_c(z)}{(1+z)H(z)}R_{\rm{TBH}}(z) \\ \times 
 C_{\theta}\left[\frac{\rho_0}{8}\frac{D_L(z)}{r_0} \left(\frac{1.2M_{\odot}}{(1+z)\mathcal{M}_c}\right)^{5/6} \right]\,,
 \end{multline}
where $D_c(z)= \int_{0}^{z}\frac{dz^{\prime}}{H(z^{\prime})}$ denotes the comoving radial distance and $D_L(z)=(1+z)\,D_c(z)$ denotes the luminosity distance respectively. $H(z)$ is the Hubble rate at redshift $z$ and we use the cosmological parameters determined by the latest {\sc Planck} observations~\cite{Aghanim:2018eyx}. The angular dependence of the signal-to-noise-ratio (SNR) is encoded within the variable $\theta$, and the cumulative distribution of $\theta$ is denoted by $C_{\theta}$~\cite{Taylor:2011fs}. The chirp mass of the coalescing binary is denoted by $\mathcal{M}_c$, and $\rho_0,r_0$ are the SNR threshold for GW detection and the characteristic distance sensitivities of the GW detectors. For this analysis, we have considered $\rho_0=8$, and $r_0=80\,(1591)$ Mpc for {\sc aLIGO} ({\sc ET})~\cite{Taylor:2012db}. One finds that already {\sc aLIGO} would be sensitive to DM parameters that are not ruled out by other data at present, e.g., $m_\chi=10^4$\,GeV and $\sigma_{\chi n}=10^{-45}$\,cm$^2$. Suppose $N_D=5916$ yr$^{-1}$ in the low redshift bin, which could be due to mergers of 1.3 $M_{\odot}$ PBH binaries with $f_{\rm{PBH}}=0.0019$ or 1.3\,$M_{\odot}$ TBH binaries with $m_{\chi} = 10^{4}$ GeV, $\sigma_{\chi n} = 10^{-45}\,\rm{cm}^2$. 1\,ET-yr worth of high redshift data can discriminate between these two model points using shape-information alone (i.e., with same normalization at low redshift): the PBH binaries will have $787$ events in the high redshift bin, whereas TBH binaries will have $880$, indicating putatively $\geq$\,3$\sigma$ discrimination between said TBH and PBH model points, with statistical errors only.

\begingroup
\squeezetable
\begin{table}
\ra{1.3}
\begin{ruledtabular}
\begin{tabular}{ccccc}
	$M_{\rm{NS}}$ $[M_{\odot}]$&$m_{\chi}$ [GeV]& $\sigma_{\chi n}$ [cm$^2$] & {\sc aLIGO} [yr$^{-1}$] &{\sc ET}  [yr$^{-1}$]  \\ \hline
	1.0& $10^4$ & $10^{-47}$ & 0.2;\,0;\, 0.2& 672;\,3;\,675   \\ 
	1.0& $10^4$ & $10^{-45}$ &  0.3;\,0;\,0.3&2982;\,32;\,3014  \\ 
	1.3& $10^4$ & $10^{-47}$ & 0.4;\,0;\,0.4&1451;\,84;\,1535 \\ 
	1.3& $10^4$ &$10^{-45}$  & 0.8;\,0;\,0.8&5916;\,880;\,6796
\end{tabular}
\end{ruledtabular}
\caption[]{Possible detection rates of TBH binaries for {\sc advanced LIGO} ({\sc aLIGO}) and {\sc Einstein Telescope} ({\sc ET}), estimated using the procedure in the text, for representative choices of NS mass $M_{\rm{NS}}$, dark matter mass $m_{\chi}$ and DM-nucleon scattering cross section $\sigma_{\chi n}$. The radius of the progenitors are taken to be 10.6 km. The three numbers in the last two columns are for low redshift ($z \leq 1$); high redshift ($z > 1$); and total, respectively.}
\label{table}
\end{table}
\endgroup

The ambient DM density around a sub-Chandrasekhar or $\mathcal{O}$(1)\,$M_\odot$ BH is a simple yet powerful probe of the origin of the BH.  Since a DM rich environment favors implosion of stellar objects, detection of a low mass BH in a DM deficient region will prefer a primordial origin.  Coexistence of a low mass BH and a NS of similar age can also be a strong evidence of its primordial origin, as the required parameter space for such transmutation will be disfavored by the existence of the companion NS.  Since the DM dense inner regions potentially contain a large number of NSs~\cite{Safdi:2018oeu}, detection of an $\sim \mathcal O (1)$ Gyr old NS by the radio telescopes like {\sc FAST}\,\cite{2011IJMPD..20..989N} and {\sc SKA}\,\cite{Bull:2018lat} will significantly strengthen the exclusion limits. As a consequence, the allowed parameter space for dark core collapse-induced transmutation of a stellar object will shrink.

Mass distributions of the compact objects provide yet another powerful way to distinguish TBHs from PBHs.  Since, the TBHs track the mass distribution of their progenitors, it can be compared against well-motivated PBH mass distributions to statistically determine the stellar or primordial origin of BHs. The last two panels of Fig.\,\ref{fig: mass distribution merger rate} correspond to the mass distribution of all observed NSs and white dwarfs, progenitors of the TBHs. It would take a striking coincidence for the PBH mass distribution to coincide with these distributions. Ref.~\cite{Takhistov:2020vxs}, that appeared as our paper was being readied, establishes this technique in more detail. 
    
With imminent ground and space-based GW detectors, about one binary NS merger event is expected per week\,\cite{Aasi:2013wya}. Considering the huge number of expected events, the greatly improved sky localization of the GW events with a multi-detector network\,\cite{Aasi:2013wya}, as well as the GW lensing\,\cite{Hannuksela:2020xor}, the implosion scenario can easily be tested in the near future. Observationally, there also exist several ways to distinguish a TBH binary from a binary NS. The peak signal frequency of a binary NS merger is much lower than that of a binary BH merger due to the less compact nature of NSs compared to the similar mass BHs\,\cite{Kouvaris:2018wnh}. The amount of ejected NS material during merger is much larger if one of the components is a BH, and therefore, an unusually bright transient would favor a
low mass TBH-NS merger~\cite{Yang:2017gfb}. Besides, the dimensionless tidal deformability parameter, which is zero for a BH and $\sim 100$ for a typical NS, and the strength of tidal heating can also be used to probe this implosion scenario\,\cite{Fasano:2020eum,Datta:2020gem}. More importantly, possible detection of an associated electromagnetic counterpart from radio wavelengths to gamma rays can also distinguish binary BHs from binary NSs or BH-NS merger.
  
 \emph{Summary \& Outlook --} Sub-Chandrasekhar mass BHs cannot be explained by stellar evolution and will herald new physics.  PBHs are the most discussed explanation of these objects. The  notable existing alternative proposals\,\cite{Capela:2013yf,Kouvaris:2018wnh,Shandera:2018xkn} are either not effective or appeal to baroque DM models.  We study a simple mechanism for transmutation of compact objects that can produce low mass BHs without fine-tuning. Non-zero interactions with stellar nuclei, which is a universal feature of DM models, is sufficient for such transmutations if annihilations are absent. For sub-Chandrasekhar mass progenitor, the imploded BH is a viable alternative to PBHs, whereas, for a heavier mass progenitor, it can also possibly explain the lighter companions of recent anomalous GW events. Cosmic evolution of the merger rate and the mass distributions of the progenitors are simple yet powerful probes of our proposal. Observation of an associated electromagnetic counterpart along with a GW event, as well as a precise measurement of the tidal deformability parameter, can differentiate merger of such TBHs from a binary NS merger or a BH-NS merger. Importantly, possible detection of any low mass BH in a DM deficient environment or accompanied by an old NS can falsify our proposal. Improved sky localization with multi-detector networks as well as sub-arc-second precision of a GW event from GW lensing can also shed light on this topic in the near future.

 \emph{Acknowledgments --} We  thank Parameswaran Ajith, Manjari Bagchi, Anumita Bose, and Lankeswar Dey for useful discussions. B.D. is supported by the Dept. of Atomic Energy (Govt.\,of India) research project under Project Identification No. RTI 4002, the Dept. of Science and Technology (Govt.\,of India) through a Swarnajayanti Fellowship, and by the Max-Planck-Gesellschaft through a Max Planck Partner Group. R.L. thanks CERN Theory group for support.

\bibliographystyle{JHEP}
\bibliography{ref.bib}

\end{document}